\documentclass[12pt,english]{article}
\usepackage{mathptmx}

\usepackage[T1]{fontenc}
\usepackage[latin9]{inputenc}
\usepackage{geometry}
\usepackage{babel}
\usepackage{calc}
\usepackage[authoryear]{natbib}
\usepackage[unicode=true,pdfusetitle,
 bookmarks=true,bookmarksnumbered=false,bookmarksopen=false,
 breaklinks=false,pdfborder={0 0 1},backref=false,colorlinks=false]
 {hyperref}

\makeatletter
 \let\oldquote\quote
      \renewcommand\quote{\small\oldquote}
      \let\oldquotation\quotation
      \renewcommand\quotation{\small\oldquotation}
\date{\vspace{-5ex}}
\providecommand{\keywords}[1]{\textbf{\textit{Keywords }} #1}
\setcitestyle{round,aysep={}}

\makeatother

\begin{document}

\title{OPEN SOURCE AND SUSTAINABILITY: THE ROLE OF UNIVERSITY}

\author{Dr. Giorgio F. SIGNORINI, PhD\\
Dipartimento di Chimica, Università di Firenze\\
via della Lastruccia, 3\\
I-50019 Sesto F. (Firenze), Italy\\
giorgio.signorini@unifi.it }
\maketitle
\begin{abstract}
One important goal in sustainability is making technologies available
to the maximum possible number of individuals, and especially to those
living in less developed areas (Goal 9 of SDG). However, the diffusion
of technical knowledge is hindered by a number of factors, among which
the Intellectual Property Rights (IPR) system plays a primary role.
While opinions about the real effect of IPRs in stimulating and disseminating
innovation differ, there is a growing number of authors arguing that
a different approach may be more effective in promoting global development.
The success of the Open Source (OS) model in the field of software
has led analysts to speculate whether this paradigm can be extended
to other fields. Key to this model are both free access to knowledge
and the right to use other people's results. 

After reviewing the main features of the OS model, we explore different
areas where it can be profitably applied, such as hardware design
and production; we finally discuss how academical institutions can
(and should) help diffusing the OS philosophy and practice. Widespread
use of OS software, fostering of research projects aimed to use and
develop OS software and hardware, the use of open education tools,
and a strong commitment to open access publishing are some of the
discussed examples.
\end{abstract}
\keywords{Open Source, Sustainable Development, University, Open
Education, Open Access} 

\section{Introduction}

What is sustainability about? According to the widely accepted definition
of the Brundtland Report \citep{brundtland1987report}, human development
is sustainable when it can satisfy the needs of the current generation
without compromising the ability of future generations to do the same.
This is the original approach, which puts the focus on \emph{resource
consumption}; for example, it is evident that using renewable sources
for the production of energy is sustainable, while consuming exhaustible
fossil fuel reserves is not. 

However, it has long been recognized that there are many aspects of
human growth, other than the depletion of natural goods, that can
be not sustainable \citep{Brandt1980north,Brandt1983common,quilligan2002brandt}:
among them, uncontrolled population growth, the rush to armaments,
an ever-rising debt of poor nations. Less obviously, other issues
such as unequal distribution of wealth or the discrimination of women
are also seen as non-sustainable, simply because they inevitably lead
to social and political instability. In the course of years, the sustainable
development objectives promoted by the United Nations, first in 1992
(``Agenda 21''), then in 2000 (``Millennium Development Goals'')
and again in 2015 (``Sustainable Development Goals'', SDG \citep{SDG}),
have come to include more and more economic and social issues. 

One of the SDGs (goal 9: ``Build resilient infrastructure, promote
sustainable industrialization and foster innovation'') directly involves
the transfer of technology (``innovation'') to less developed countries.
It is hard to overlook the striking contrast that currently exists
between the high level of technology reached by humanity as a whole
and the large fraction of people having no access to it \citep{pearce-2012-OSAT};
think of life-saving drugs which major pharmaceutical companies hold
the patents of, or of the technical/scientific literature that is
only published on expensive journals most educational institutions
in the Third World cannot afford to buy. Indeed, the lack of access
to, and command of, technology has been described sometimes as the
\emph{main} weakness of developing countries (\citep{Brandt1980north},
cited in \citep{quilligan2002brandt}).

It is a fact that one of the obstacles, perhaps the most effective
one, to the diffusion of technology is represented by the regulations
protecting the so-called Intellectual Property Rights (IPR). Opinions
about how effective IPRs are in promoting and disseminating technical
innovation differ (see, for example, \citep{blind-2012-regulations}
and references therein). The traditional view has been that IPRs are
required in order to secure a form of reward for the research investment.
However, in recent years there has been a growing number of studies
pointing out that a different paradigm may be more effective in fostering
innovation \citep{Weber-2000-economy_of_OSS,boldrin2009against,henry_and_stiglitz-2010-IP_and_innovatn,boldrin-2013-against_pat,daley-2014-GSDR-optimality,stiglitz-2015-learning}.

There are two main ways IPRs can hinder development of poorer nations:
by limiting people's access to knowledge through copyright, and by
restricting the use of novel technologies through patents. Thus, an
alternative model should be able to address both issues. 

What is commonly known as Open Source Software (but is more appropriately
termed FLOSS, see below), has challenged the current production paradigm
in the area of information technology by explicitly tackling these
two aspects. Note that the expression ``Open Source Software'' (OSS),
in fact, only implies removing the first of the two obstacles, regarding
\emph{availability}; however, in the general parlance, it also encompasses
the \emph{right to use} the accessed resource. The success that OSS
has seen in recent years \citep{bonaccorsi2003open} has led many
authors to suggest that the Open Source (OS) scheme be exported to
other areas, such as hard technologies, to favor their advancement.

``Open Source Hardware'' \citep{OSHWA} and ``Open Design'' initiatives
are attempts in this direction, that have contributed both the theoretical
framework of the approach and concrete examples of how it can be implemented
and sustained \citep{li2017open}. The ``Open Access'' movement
advocates free access to (and use of) any kind of intellectual work,
including the scientific and technological literature, which can encourage
innovation in less developed countries. Educational institutions are
increasingly investing in ``Open Education'' programs. Despite the
rich literature that exists on these subjects, few authors have tried
to discuss in a unified, comprehensive fashion the concept of ``openness''
in such different contexts \citep{Pomerantz-2016-Fifty,Aksulu-2010-review_OS}.
The present paper is an attempt to fill in this gap, with special
emphasis on what elements of the OSS model are retained in each, and
on their implications with respect to sustainable development.

A second point of this work derives from the observation that key
to the diffusion of the OS philosophy to new areas is how it is perceived
by the public. OSS has been traditionally viewed by the general opinion
as only a cheap alternative to quality products, but recently this
perception is changing, with a growing interest in OSS by companies
and public administrations \citep{roumani2017trust,Casson-2006-OS_in_public}.
Obviously the education and research world plays a primary role in
this change of perspective \citep{Coppola-2004-OS_in_edu,Futurelab-2006-education,Lakhan-2008-OSS_in_educ,OHara-2003-OSS_in_educ,Pankaja-2013-OSS_in_edu},
because they can not only illustrate the advantages of these products
or the ethical motivations that are at the roots of the philosophy,
but also support OS with working examples. The effort that universities
will be able to put in this field can probably make the difference.

In the following, a review of the main features of OSS is presented
first; then, in the central section, a number of possible areas of
application of the ``Open Source'' model are analyzed; finally,
we discuss the role of university in promoting the diffusion of the
Open Source model.

\section{The lesson of Open Source Software}

\subsection{``Open Source'' vs. ``Free'' (or, use FLOSS regularly)}

As anticipated above, although the designation ``Open Source Software''
has gained widespread acceptance by now, it is very misleading. In
almost all contexts, in fact, it is applied to software that can be
not only accessed freely but also legally used and distributed; that
is, basically, what the early (circa 1985) definition of ``Free Software''
by Richard Stallman and the Free Software Foundation (FSF) \citep{FreeSW}
established:

\medskip{}

\noindent %
\noindent\fbox{\begin{minipage}[c]{1\textwidth - 2\fboxsep - 2\fboxrule}%
\begin{quote}
A program is free software if the program's users have the four essential
freedoms:

\begin{itemize}
\item The freedom to run the program as you wish, for any purpose (freedom
0). 
\item The freedom to study how the program works, and change it so it does
your computing as you wish (freedom 1). Access to the source code
is a precondition for this. 
\item The freedom to redistribute copies so you can help others (freedom
2). 
\item The freedom to distribute copies of your modified versions to others
(freedom 3). By doing this you can give the whole community a chance
to benefit from your changes. Access to the source code is a precondition
for this. 
\end{itemize}
\end{quote}
\end{minipage}}

\medskip{}

Note that here the stress is placed on freedom, the rights that are
granted to the user. While freedom 0 may sound rather obvious, in
analogy with what one is entitled to do with any device they may obtain,
freedom 1 is a little more problematic. Think of a buying a small
appliance: you surely have the right to inspect and possibly modify
it, but in practice you can't, because the operation of an increasing
number of them relies on instructions coded in electronics, which
are difficult if not impossible to understand. This is where the concept
of openness comes in: openness is a necessary requirement to enable
the user to fully control their device. Freedoms 2 and 3 give the
user the right of reproducing the item, something that is usually
not permitted with real objects, at least with those covered by patents.

On the other hand, it is to be remarked that nothing about the costs
(``think of \textquoteleft free speech\textquoteright , not \textquoteleft free
beer\textquoteright '' \citep{FreeSW}) is implied by the above definition;
this kind of software can be profitably traded - in just about the
same way that a bottle of water from a mountain creek can. To disambiguate
between the two different meanings of the English word ``free'',
the terms ``gratis'', as opposed to ``libre'', are sometimes used.

In spite of these important semantic distinctions, ``Open Source''
has now come to assume a much broader meaning than the words encompass,
especially so after the founding (1998) of the Open Source Initiative
(OSI). OSI's now widely recognized definition of Open Source Software
\citep{perens-1999-OS_def,OS-def-OSI} closely resembles the one by
the FSF:

\medskip{}

\noindent %
\noindent\fbox{\begin{minipage}[c]{1\textwidth - 2\fboxsep - 2\fboxrule}%
\begin{quote}
Generally, Open Source software is software that can be freely accessed,
used, changed, and shared (in modified or unmodified form) by anyone
\citep{OS-def-FAQ}
\end{quote}
\end{minipage}}

\medskip{}

There are still some fine differences between FSF, OSI, and other
definitions, which however are not relevant for our purpose. Perhaps
the best designation for this technology is the \emph{portmanteau}
``Free (Libre), Open Source Software'' (FLOSS), which, if somewhat
redundant, effectively transmits the notions of both freedom and openness.

\subsubsection{copyleft}

The legal frame for distributing FLOSS is a set of licenses that protect
the basic freedoms of the user. The one that FSF propose and use for
their software, such as the GNU suite which is an essential part of
the GNU/Linux operating system, is the so-called ``copyleft'' (where
the second half of the word contrasts the one in ``copyright''):
copylefted software is subject to distribution terms that ensure that
copies of that software carry the same distribution terms. The license
that formally details these terms is the GNU General Public License,
or GPL.

\subsection{Quality}

The first issue regarding FLOSS is about its general quality.

There is a widespread view that ``since Open Source software is free,
it must be of low quality''. This idea is deeply rooted in our everyday
experience: quality goods have high prices and their ownership is
strictly protected.

However, software is a good of complex nature, as noted by many authors
(\citep{bonaccorsi2003open,Weber-2004-OSsuccess}). In fact, it turns
out that in many cases the performance of FLOSS is comparable or superior
to that of their proprietary counterparts. Studies of the last two
decades have shown many FLOSS products to be highly \textbf{reliable}
(in the sense of both ``stable'' and ``secure''), and in many
cases to outperform proprietary systems \citep{boulanger2005open}.
For example, in a well-known test, commercial Unix systems had a failure
rate that ranged from 15\% to 43\%; in contrast, the failure rate
for GNU was only 7\% \citep{miller1995fuzz}. Another feature that
adds to the quality of open-source software is its high degree of
\textbf{flexibility}, which means both that it can be easily customized
\citep{lerner2004economic} to meet different or new needs, and that
it can be very resilient to changes in the environment. 

While it may be not easy to precisely define the ``quality'' of
software products, there are some valuable -if indirect- measures
of it: for example, the level of diffusion of OSS, and the motivations
that drive its adoption. Of course, data referring to organizations
and companies are more desirable than those about individuals, as
the former are expected to base their decisions on an objective evaluation
of advantages and disadvantages, rather than on personal preferences
and opinions.

There are not many surveys of general scope regarding the popularity
of OSS; most data deal with network applications, which can be easily
monitored. It is known, for example, that among web server programs
and the underlying operating systems, FLOSS usually ranks first \citep{Wheeler-2007-whyOSS}.
Results vary considerably and depend, among other things, on country,
activity sector, size of organization \citep{Ghosh-2002-OS_survey,VIU-2009-OSinItalia,Wheeler-2007-whyOSS};
however, the fact that open source solutions occupy a significant
share of the market, especially in the field of server systems, is
universally recognized, as is the fact that their popularity is constantly
rising.

``Quality of solutions'' and ``competitive features and technical
capabilities'' are cited in a recent survey \citep{duck-2016-survey}
as the first two reasons why experts adopt open source. According
to another study \citep{roumani2017trust} the three main sources
of trust in enterprise-OSS are: conformation to open standards; security;
service. Rather unexpectedly, in almost all reports cost does \emph{not}
emerge as the main motivation behind the choice of adopting FLOSS.
In fact, open products prove cheaper than proprietary ones, in general,
only if the total cost of ownership (TCO), rather than the sheer cost
of adoption, is considered. Users are preferably attracted by other
positive features, such as stability, security, user experience, compatibility,
transparency, customizability \citep{GitHubOpenSourceSurvey2017},
and also the availability of service \citep{Benkler-2006-networks}.
We will discuss these aspects in the next section.

As a whole, we can safely conclude that there are many FLOSS products
that are of very good quality, although this is, obviously, not automatically
true of \emph{all} FLOSS.

\subsection{Features}

According to the Open Source Initiative \citep{OS-mission-OSI}, 
\begin{quote}
``Open source enables a development method for software that harnesses
the power of distributed peer review and transparency of process''
\end{quote}
In the following we list and shortly comment on the distinctive features
of FLOSS that make this possible. All these features are direct consequences
of the basic properties that define FLOSS (in either the FSF or the
OSI version, see above), and ultimately of the two basic rights: the
right to \emph{access }and the right to \emph{actively use} it.

\subsubsection{reliability}

By design, in an open source project there is no limit to the number
of contributors, with every user being a possible developer, and popular
projects involving thousands of them \citep{openhub}. It is a now
generally accepted view that a large community performing the revision
and test process provides fast and efficient bug fixing, vulnerability
checking, performance refinement; as early as in 1999, Eric Raymond
in his seminal essay \emph{The Cathedral and the Bazaar} \citep{raymond1999cathedral}
was boiling this concept down to 
\begin{quote}
``Treating your users as co-developers is your least-hassle route
to rapid code improvement and effective debugging''
\end{quote}

\subsubsection{flexibility}

The diversity of the environments where open source programs are developed
and used, and the fact that most people that support FLOSS are both
users and developers is also at the roots of its great flexibility
\citep{roumani2017trust}. Localization, implementing of new features,
adapting tools to changed conditions: all these tasks are more easily
carried out by a sharing community than by a small number of hired
experts who must respect the secrecy and patent restrictions as is
typical of commercial software firms.

\subsubsection{innovation and learning incentive}

It has been noted that the open source model also has a greater potential
for innovation (e.g., for filling unfilled market areas \citep{boulanger2005open}).
New ideas are best fostered in a free and knowledge-sharing environment.
It is a fact that many of the tools that have made the revolution
of the ICT world in the last decades, such as Internet and Internet
applications like electronic mail and the WWW, the Android-based smartphones,
and Wikipedia, were based -if not on open-source software in a strict
sense- on open standards and/or shared technologies. Using Raymond's
words again \citep{raymond1999cathedral},
\begin{quote}
``the root problem of innovation (in software, or anywhere else)
is {[}...{]} fundamentally {[}...{]} how to grow lots of people who
can have insights in the first place.''
\end{quote}
that is, it is by reaching widest possible diffusion of knowledge,
and not by restricting it with IPRs, that we can favor innovation.

The ability of reading the program codes and modifying them at will,
opens better learning opportunities for users of FLOSS. This is why
FLOSS is thought to be an ideal candidate for the use in educational
environments \citep{Futurelab-2006-education}.

\subsubsection{collaborative scheme}

Many FLOSS advocates claim that the main value of it lies in the production
method itself. Setting up work in a way that is ``radically decentralized,
collaborative, and nonproprietary; based on sharing resources and
outputs among widely distributed, loosely connected individuals who
cooperate with each other'' \citep{Benkler-2006-networks} is indeed
a radical change from the traditional, hierarchically organized and
competition-driven, perspective. Some authors, both in the economics
and in the educational field, have long questioned the validity of
the widely accepted assumption that setting people ``one against
the other'' is the best way for boosting production \textendash let
alone living happily \citep{kohn-1992-nocontest,kohn-1987-reward_no_motiv,stiglitz-2015-learning}.
FLOSS provides a real-world example of how work can be organized in
a totally different way from the traditional one and still be as efficient
\textendash or even more. 

Thus one of the advantages of using FLOSS is of a social nature: 7it
promotes changes in society that may help build a more sustainable
world. 

\subsubsection{independence from vendor}

From the user point of view, proprietary software often has the undesirable
effect of forcing clients to keep using the same software even when
it no longer meets their original needs. This is due to the use of
proprietary formats or tools that cannot be exported to a different
platform, perhaps as the result of an aggressive fidelization policy
of the vendor.

FLOSS has no blind spots. Migration to a new software is always possible
because users have full control of algorithms and data. In some cases
this process may be painful, but it is likely that the community will
come to help with compatibility and conversion software. And this
will be so forever, while a discontinued proprietary software may
result in your resources becoming unusable with its secret machinery
buried in some unaccessible archive, or lost for good.

Freedom from vendor lock-in \citep{roumani2017trust} is especially
desirable in the public sector \citep{Casson-2006-OS_in_public}.

At the opposite end of closed technology lies the system of \textbf{open
standards}. FLOSS spontaneously encourages the formation of standards
\citep{Weber-2000-economy_of_OSS}, which can have beneficial economical
effects. Firms may choose to adopt FLOSS to help ``development of
open standards and discourage establishment of a proprietary one''
\citep{bonaccorsi2003open} by some competitor.

\subsubsection{low cost}

FLOSS can be distributed at lower prices than commercial products,
as a consequence of reduced costs of both production and marketing.
As already pointed out, FLOSS does not necessarily come at no cost:
storage media for recording the program, shipping, and the like, do
require some expense, and can be conveniently provided by some distributor
\textendash which, as a side-effect, opens a market opportunity for
new initiatives. 

As we have seen, low cost does not represent the main motivation in
the adoption of FLOSS by companies or professionals; however, it can
have nonnegligible, beneficial effects on their budgets. 

\subsubsection{service}

Support services are perceived by companies as an essential requirement
of software products \citep{Benkler-2006-networks}. On the one hand,
often proprietary software suppliers also offer support contracts
(usually reliable), while FLOSS distributors do not necessarily have
the expertise to provide that service. On the other hand, again, since
there is no restriction on studying FLOSS, it can be potentially serviced
by anyone \textendash especially the developers themselves. Actually
this represents a very good opportunity for ``the emergence of local
capabilities to provide software services'' \citep{Benkler-2006-networks}

\subsection{Examples}

Although this is not intended as a comprehensive catalog of FLOSS,
it is nonetheless useful to mention a few products that are especially
popular:
\begin{itemize}
\item \textbf{GNU/Linux} is, in a way, the archetypal free software. In
the 1990s, given the complex license status of the UNIX operating
system, the need of a free system led, on one side, to the creation
of the kernel (Linux), and on the other side to the creation of the
set of basic utilities that make the user interface (GNU). 
\item \textbf{Android}, the leader operating system of smartphones (77\%
market share in 2018 \citep{statcounter}) and now also the first
operating system worldwide (42\% market share \citep{statcounter}),
is based on a Linux kernel but according to somebody cannot be properly
called FLOSS since it is released under a license that is more restrictive
than copyleft.
\item among server applications, the \textbf{Apache HTTP server} (also not
properly FLOSS) together with Nginx makes 84\% of active web servers
\citep{w3techs}
\item among end-user applications, \textbf{Mozilla Firefox} web browser,
once the most popular web browser, is now (2018) the second most popular
one on desktop/laptop computers after Chrome, with an 11\% market
share \citep{statcounter}; 
\item where do you look up for general reference? There's more than a chance
it's \textbf{Wikipedia}, ``one of the most successful collaborative
enterprises that has developed in the first five years of the twenty-first
century'' \citep{Benkler-2006-networks}. Wikipedia is relevant in
this context for multiple reasons: it is operated by a FLOSS called
MediaWiki and is in itself one of the best examples of the validity
of the core features of the OS model. It is open to anyone to contribute;
every content including older versions of the articles is freely accessible;
and it is universally considered as highly reliable \citep{Benkler-2006-networks}
\item it is appropriate, for its importance, to list here the \textbf{TCP/IP}
protocol, the standard on which Internet is based on: Although not
properly a software, Internet has established itself over the existing,
various computer network technologies (or software) because it was
designed to be independent of proprietary specifications \textendash a
key feature it shares with FLOSS.
\end{itemize}

\section{The OS model and its possible applications}

Perhaps the best way to summarize the above discussion in view of
the first point of this paper is the following sentence:
\begin{quote}
``Now that Open Source has come of age, the question is not: Is it
better than closed software? But rather: To what other systems, outside
of software, can we apply the concepts of Open Source and public ownership?''
\citep{Agroinnovation}
\end{quote}
More specifically, we want to ask ourselves:
\begin{itemize}
\item can the OS model be exported to \emph{hard} technologies? and perhaps,
in a broader sense, to the worlds of content publishing and education?
\item which of the defining properties of FLOSS can also be applied to these
areas?
\item what are the differences?
\end{itemize}
Note that key to any OS project are: a network infrastructure through
which contributors can share their work and ideas; a sound system
of governance that effectively channels activity into the target product;
a software platform that implements such a system \citep{bonaccorsi2003open}.

\subsection{Open Source Hardware}

There is one obvious difference between hardware and software production:
software, as opposed to hardware, is immaterial. To obtain and use
a computer program everything that is needed is some digital storage,
a very cheap resource; on the contrary, the building of a technical
equipment always requires a certain amount of starting materials that
must be supplied by the user, generally at non-negligible costs. 

Clearly, open-source hardware is not about sharing the ownership of
physical devices or tools; it is about free access only to their \emph{immaterial}
part, namely blueprints, methods, all the know-how needed. As we have
seen, the necessary and sufficient conditions of the OS model are
knowledge sharing and the right to actively use it. 

The idea of ``Open Design'' is basically that of directly projecting
the principles of FLOSS onto the world of machinery and manufacturing
processes \citep{Vallance-2001-opendesign}, as an alternative to
the proprietary design scheme and with the same motivations as open-source
software: favoring innovation, quality and accessibility of products
through collaboration of experts and users alike. The Open Design
Foundation has established an ``Open Design Definition'' \citep{ODF}
and terms of use that closely resemble those of FSF \citep{Vallance-2001-opendesign}:

\medskip{}

\noindent %
\noindent\fbox{\begin{minipage}[c]{1\textwidth - 2\fboxsep - 2\fboxrule}%
\begin{itemize}
\item documentation of a design is available for \emph{free,}
\item anyone is \emph{free} to use or modify the design by changing the
design documentation,
\item anyone is \emph{free} to distribute the original or modified designs
(for fee or for free), and
\item modifications to the design must be returned to the community (if
redistributed).
\end{itemize}
\end{minipage}}

\noindent \medskip{}

\noindent Similar ``Open Source Hardware'' definition and principles
have been provided by the Open Source HardWare Association (OSHWA)
\citep{OSHWA}

\subsubsection{Open Source Appropriate Technologies}

Experience of the last decades has shown that the ideas of OS can
be successfully applied to the area of the so-called ``Appropriate
Technologies'' (AT). Originally proposed \citep{Schumacher-1973-small}
as a response to the evidence that cutting-edge, sophisticated technologies
produced by the highly developed nations usually promote very little
advance in the quality of life of the majority of world population
(low income classes and/or countries), ATs can be defined as ``technologies
that are easily and economically utilized from readily available resources
by local communities to meet their needs'' \citep{pearce-2012-OSAT};
they are ``appropriate'' with respect to the regional size of the
economy which they are supposed to help \citep{Schumacher-1973-small}
and which is seen as the ideal size for the development of those parts
of society who need it most.

As it has been pointed out, ``more than 10 million children under
the age of five die each year from preventable causes'', in spite
of the cures being well known, just because they are not available
economically \citep{pearce-2012-OSAT}.

Small size and limited complexity are key features of ATs. However,
some technologies, even very basic ones, may still be inaccessible
to under-privileged communities not so much because of their cost,
as because of the lack of the knowledge needed to use and maintain
them: think of a patented device which may be operated only by skilled
professionals licensed by the manufacturer \citep{mushtaq-2011-OSAT}.
Open Source Appropriate Technology (OSAT), patent-free AT which everybody
may copy the design of, fits well in this scenario, and is aimed at
cutting monopoly/royalty costs while giving users-developers full
control over their equipment (for example, allowing them to incorporate
a small apparatus in a larger one).

The infrastructure supporting OSATs usually takes the form of an open
clearinghouse, like Appropedia \citep{appropedia}, storing all the
instructions for building the solutions proposed. Not surprisingly,
Appropedia is based on MediaWiki software (see above) and its content
is licensed under a Creative Commons BY-SA license (see below). 

\subsubsection{Manufacturing}

As the Open Design movement shows \citep{Vallance-2001-opendesign,li2017open},
open-source machinery does not necessarily have to be simple. With
automated manufacturing, rather complex devices can be assembled,
using publicly available instruction sets. A big step forward in this
field was made since the appearance of 3D printers on the market. 

The RepRap (Replicating Rapid Prototyping) project \citep{Jones-2011-reprap},
for example, is based on one 3D printer that can (almost) replicate
itself, being able to print the majority of its own parts, and is
intended as an open source, low-cost manufacturing machine. In principle,
such machines should enable any individual to autonomously build e.g.
many of the artifacts used in an average household, on a path of increasing
independence of people from large-scale manufacturing corporations.
The potential impact of this process on global economy is evident.

Among many similar initiatives, one of more general scope is Open
Source Ecology \citep{stokstad-2011-OSEcology}, whose declared goal
is 
\begin{quote}
to create an open source economy \textendash{} an efficient economy
which increases innovation by open collaboration
\end{quote}
OSE flagship project is the Global Village Construction Set \citep{globalvillage},
a collection of open-source instructions for building what they think
is the minimum set of 50 tools needed by ``an entire self-sustaining
village'': from a tractor to an oven, to a circuit maker, to power
production stations. These include fabrication and automated machines
that make other machines \textendash an analogue of 3D printers with
a broader purpose.

\subsubsection{Examples}

As a representative example of Open Source Hardware we may take the
Arduino board \citep{arduino,Badamasi-2014-arduino}. Arduino is a
line of open-source electronic platforms with micro-controller for
the remote control of devices. 

A wealth of Open Design / OSAT projects making use of Arduino have
been implemented. We encourage readers to visit any OSH clearinghouse
to appreciate the diversity of applications this modular hardware
can be adapted to.

Arduino can also be taken as a working example of how OSH can be profitable;
this is the subject of next section.

\subsubsection{Business model}

It is natural to ask ourselves how developers of OSH can make a profit
from their work, given the lack of IPR-related revenues. 

First of all (and differently from FLOSS) the physical realization
of products accounts for an important share of hardware business,
as does its marketing. The design phase, on the contrary, is usually
not the main activity, so it is not not strictly necessary for it
to be profitable in itself. Consider that there is no sharp boundary
between developers and manufacturers: companies and professionals
often play both roles at the same time.

Secondly, manufacturers of OSH have several competitive advantages:
they don't have to pay for patents; they get an efficient user feedback
for free; their services like customer care or localization are highly
valued, since, as active developers, they know their product well
\citep{thompson2011build}. 

Arduino inventors and original makers, who decided to put its design
in the open for everybody to read, find that their product is still
more requested than the cheaper models manufactured by factories all
over the world using the open blueprints \citep{thompson2011build}.
This is because Arduino's original items turn out to be higher quality;
and since the development process, that has their company as the primary
hub, is always in progress, they find themselves always one generation
ahead of their (non-)competitors.

\subsection{Open Access}

We see that the patents system acts as a barrier to free access to
knowledge, and consequently to the diffusion of technology to less-developed
areas of the world. The same effect is caused by another class of
restrictions, namely the copyright laws.

In response to copyright, an Open Access philosophy has emerged. As
the 2002 manifesto of a group of leading academics, chief librarians,
and information officers puts it, ``Removing access barriers to ...
literature will accelerate research, enrich education, share the learning
of the rich with the poor and the poor with the rich, make this literature
as useful as it can be, and lay the foundation for uniting humanity
in a common intellectual conversation and quest for knowledge.''
\citep{BudapestOA}

\subsubsection{Scientific literature}

Let us focus on academic publications first. It is a fact that people
outside the research institutions have virtually no access to up-to-date
technical and scientific literature covered by copyright \textendash even
when they do not involve patents. Journal and books presenting new
ideas and discoveries, essential for stimulating innovation, usually
come at a forbiddingly high price for an individual or medium-sized
business. Expenditures of research libraries for bibliographic materials
have been steadily increasing in the last years \citep{ARLstat}.
Indeed, a fairly large number of universities have long declared that
they can no more afford to buy journal subscriptions \citep{sample2012harvard}.
Robert Darnton, the past director of Harvard Library, once declared
in an interview: \textquotedbl{}We faculty do the research, write
the papers, referee papers by other researchers, serve on editorial
boards, all of it for free \dots{} and then we buy back the results
of our labor at outrageous prices.\textquotedbl{} \citep{sample2012harvard}

The reason for the hyper-inflation of scientific journals is simple:
publishers operate in a basically monopolistic regime \citep{shieber2009equity,bjork2014developing}
and can impose whatever price they set. Moreover, the academic career
system, based on publications, virtually obliges scholars to publish
at any cost, thus consolidating the monopoly.

There is also the important question of whether it is fair to restrict
access to the results of research that is publicly funded. This amounts
to using taxpayers' money to subsidize a monopoly \citep{boldrin-2008-against_book}.
The U.S. National Institutes for Health (NIH) and its Canadian analogue
CIHR \citep{mushtaq-2011-OSAT} have reacted to this by requiring
that the results of the research they fund be made available to the
public. 

Open access to publications is emerging as a solution to this. The
rationale is that the cost of publishing can be payed by the authors
in order to make their articles readable for free \citep{shieber2009equity},
a mechanism that can easily be imagined to lower the overall costs
per publication \citep{Odlyzko-1997-OAjourn,BudapestOA}. However,
the Open Access (OA) journals market system has drawbacks too. Article
processing charges (APC) are still rather high, of the order of $1500\$$
per article \citep{shieber2009equity}, which sounds as a comparatively
large proportion of the total costs of a research project. Many of
the major journals, instead of switching to OA completely, maintain
a hybrid regime, both OA and subscription based, so that in the end
there is no significant reduction of costs for research institutions.
We have witnessed the birth of ``predatory publishers'', that leverage
on the researchers' need to have their articles published, but are
of very low quality and often border on fraudulent behavior \citep{pisanski2017predatory}.
This is an area of ongoing evolution and it may be still too early
to assess the efficacy of OA publishing.

There is a number of spontaneous initiatives aimed at contrasting
the current obstacles to free access to scientific literature. Many
scientists, for example, are familiar with Sci-Hub \citep{Bohannon-2016-scihub},
a platform created by Kazakhstani student Alexandra Elbakyan who strived
to get over the paywalls to the papers she needed to complete her
thesis.

\subsubsection{Alternative to copyright}

One can argue that open access not only to scientific and technical
literature, but to \emph{all }kind of content, including literary
and artistic creations, is beneficial to development. Indeed, there
is evidence that, while copyright laws limit the diffusion of intellectual
work, on the other side they have not had the alleged effect of increasing
the production of books and music \citep{boldrin-2008-against_book}.
Whether a wider diffusion of cultural products can contribute to human
development is, certainly, a debatable subject, and one that is beyond
the scope of this article. It is, nonetheless, worth noting that the
dissemination of culture -be it an invention, a painting or a novel-
is strictly connected to freedom and human rights, and thus, ultimately,
to the advancement of society. 

The alternative to property is the Commons. Several formal, legal
schemes have been devised as alternatives to the traditional copyright
model, the most prominent being the Creative Commons (CC) licenses.
These modular licenses allow authors to reserve \emph{some} (as opposed
to \emph{all}) rights for themselves, such as the moral right to be
recognized as the original author of the work. Together, the CC licenses
make up the legal frame in which cultural works can be safely distributed
while still being protected against unlawful appropriation by others
(individuals or companies).

At the opposite end of the copyright regime there are Free Cultural
Works, and the reader will not be surprised, by now, to learn that
a formal definition of FCW has been issued \citep{moller2008definition}
by the organization ``Freedom Defined'', which almost exactly matches
the FSF definition of Free Software:

\medskip{}

\noindent %
\noindent\fbox{\begin{minipage}[c]{1\textwidth - 2\fboxsep - 2\fboxrule}%
by freedom we mean:
\begin{itemize}
\item the freedom to use the work and enjoy the benefits of using it
\item the freedom to study the work and to apply knowledge acquired from
it 
\item the freedom to make and redistribute copies, in whole or in part,
of the information or expression 
\item the freedom to make changes and improvements, and to distribute derivative
works
\end{itemize}
\end{minipage}}

\noindent \medskip{}

Note that not all CC licenses fall into this definition. Namely, the
``non-commercial'' and ``no-derivatives'' clauses of CC are more
restricting than this \citep{hagedorn2011creative}. The free content
movement contends that imposing a non-commercial use license on one's
work is ``very rarely justifiable on economic or ideological grounds''
since it ``excludes many people, from free content communities to
small scale commercial users'', while ``the decision to give away
your work for free already eliminates most large scale commercial
uses''; and that those authors who want to promote widespread use
of their content should instead use a ``share-alike''-type license
like Wikipedia \citep{moller-2007ff-against_CC_NC}

\subsection{Open Education}

The open-source philosophy is relevant to the educational world under
different aspects \citep{Carmichael-2002-OS_in_edu,OHara-2003-OSS_in_educ,Lakhan-2008-OSS_in_educ}.

Openness is, obviously, at the very core of the learning process:
learning is about exploring things freely, looking at how they work,
and perhaps disassemble and assemble them again. But not everything
is as open as it seems, in the educational world. The technological
development of the last decades, largely based on non-free platforms,
has led us to accept as natural for the tools used in teaching to
be patented or copyrighted. While this seems reasonable for e.g. some
courses which need sophisticated instruments, the wide use, in schools,
of proprietary software/hardware for which there are open-source equivalents
is highly questionable. In fact, the scholar system should, in principle,
help students acquire universal skills rather than become familiar
with one particular product (and likely, be a future paying user of
it) . 

Thus, one primary issue for proponents of what can be called ``Open
Education'' (OE) is \textbf{supporting the use of non-proprietary
tools in education} \textendash not so much for reducing costs, as
for asserting \textbf{school neutrality} with respect to the market.

Also, one of the key features of the Open Source model can be profitably
applied to the learning process, namely its acting as an \textbf{incentive
for collaboration}. Since anyone can join in the modification of OS
tools, teaching projects that make use of them are very easy to implement
\citep{pearce-2007-teaching,arduino}; students can, for example,
develop a new functionality of some software or device, or add new
content to some shared cultural work (think of Wikipedia), working
together as well as with people at the other end of the world (one
example of this is the ``Google Summer of Code'' initiative \citep{summerofcode}).
Similarly, use of OS tools in the classroom can make it much easier
for \textbf{teachers to share their experiences} \citep{Futurelab-2006-education}.

From a slightly different perspective, teaching material can be made
Open Access and shared on the Internet, possibly following a wiki-like
scheme. ``Open courseware'' was initially intended as a supplement
rather than as a substitute of traditional course material, and its
parallel with open source software was explicit \citep{Long-2002-opencourseware}.
There are now a number of platforms for creating and using OA courseware,
or, more generally, \textbf{Open Education Resources} (OER) \citep{UNESCO-2015-OER};
see for example the Open Education Consortium \citep{oeconsortium},
Open Education Europa \citep{oeeuropa}, the OER Commons \citep{oercommons},
and Opensource.com \citep{opensourcecom}. Advantages of OER are -once
again- a more efficient use of resources (teachers can translate or
adapt other teachers' material without having to start from scratch
or face the copyright limitations), reduced costs for students, and
the possibility of expanding the subjects touched in the classroom
with a library of supplementary materials. It is evident how valuable
OERs can be in promoting less-favored populations' access to knowledge. 

An important class of OERs are Massive Open Online Courses (MOOC),
courses that can be completely administered through the Internet and
enable institutions to reach a much larger audience than traditional
classroom courses. A group of high-class university partners led by
MIT, Harvard and Berkeley, is giving life to the edX Consortium, in
their own words a ``MOOC provider that is both nonprofit and open
source'' \citep{edX}.

\section{The role of university}

The second point of this paper is about how universities can help
promote the Open Source model in its many different aspects and applications,
thus contributing to global sustainability. In the preceding section,
dealing with Open Education, several ways of interaction between the
scholar system and Open Source have been outlined; in the following
we illustrate, in a schematic way, the lines of action that Universities
can adopt in order to help OS gain weight and attention.

\subsection{support the Open Source philosophy}

It has been noted that the public discourse accompanying OSS can influence
the diffusion of this technology \citep{Marsan-2012-adoption}. Clearly,
besides IT specialists, higher education institutions have a primary
role in transmitting a positive or negative attitude towards OSS. 

Moreover, as we have seen, OS is not simply a new technology, but
rather a new way of producing and sharing technology and knowledge
in general. In this respect, universities can be even more influential.
By taking a clear stance in favor of the OS vision they will contribute
to advancing a positive perception of OS against the common wisdom
that it means ``no cost\textendash no quality''. 

There are many opportunities for academic institutions to underline
the positive features of the OS approach we have outlined above (reliability;
flexibility; incentive for collaboration, learning and innovation;
independence from vendor; low cost). This is best done by stimulating
the debate on the subject, through seminars, conferences and open
discussion groups. Specific courses on OS topics can help them gain
official recognition.

There are many examples of single universities or associations of
universities committed to Open Source (see for example \citep{axelerant}). 

\subsection{use Open Source tools for teaching and research}

Universities supporting the OS model can do more than just promote
it. They can actively support the use of OS software and hardware
both in teaching and research projects. There is currently a rich
literature on the subject and also a great deal of online resources
(see, e.g., \citep{Futurelab-2006-education}) that can be used for
inspiration.

The first way in which the OS approach can be put in action is by
using methods of network-based collaboration, a scheme that is central
to the OS model. Many educators, well before the advent of digital
networks, have argued that by peer-reviewing the work of one another,
students can achieve a higher level of knowledge than in the traditional,
top-down approach \citep{Scardamalia-1991-knowledgebuild}. As illustrated
in the section ``Open Education'' above, this can be accomplished
by setting up projects that involve designing new OS software and/or
hardware, or modifying existing items, with the help of collaboration
environments that should be OSS themselves.

Teaching material from such projects can be shared using existing
OE platforms (www sites, wikis, repositories), or a new platform can
be created \emph{ex novo}.

As for research work, it is common practice to carry it out in teams;
universities can set up incentives for research projects where the
subject is OS software or hardware.

\subsection{Open Access publication}

Universities are the primary sources of advanced knowledge. The majority
of them are publicly funded, either directly or indirectly. In accordance
with their mission, many of them are already committed to making as
much as possible of what they produce available worldwide under an
open access license. This can take the form of explicit rules requiring
research funded by a university to be published open access. 

Faculty and research staff are well aware that their work is already
paid for by their salaries and grants and need not be further remunerated
by copyright; and that, on the other side, free circulation of the
material they produce will help them gain visibility and reputation.

Publication costs, in the form of the APC imposed by major publishers,
are largely unjustified, given the high profits of these companies
\citep{guardian2017}. Universities and public libraries have been
(partly) successful in negotiating with publishers more reasonable
deals on APC and subscription prices, but only when negotiation is
led by a group of representative and influential institutions of a
whole country \citep{Vogel-2017-projekt_DEAL}. 

\subsection{substitute proprietary software with FLOSS}

There is a simple step universities can take toward the diffusion
of FLOSS: they can adopt the policy of substituting proprietary software
with FLOSS.

This is more effectively done in the teaching and research areas,
where the attitude towards FLOSS is usually more positive, and contact
with this kind of tools more likely to have already happened. However,
the transition to FLOSS can be made in every branch of activity of
a university, including administration and technical services, for
the good reasons that are valid for any generic firm \citep{roumani2017trust}
and especially in the public sector \citep{Casson-2006-OS_in_public},
and that have been outlined in the section ``Features'' above: reliability,
flexibility, independence from vendor lock-in, low cost.

\section{Conclusions}

The starting point of this paper is the well established view that
for our world to be sustainable (a) a substantial effort from developed
countries to help poorer ones is necessary; and (b) this help is best
given by transferring capabilities, rather than goods, so that in
the future disadvantaged communities will be able to provide those
goods for themselves. 

The extraordinary advances of the last decades in the ICT area have
made the transfer of knowledge incredibly faster and more efficient
than ever; but, at the same time, the economical and legal barriers
to free flow of information have become stronger, and tend to maintain
the current imbalance between developed and undeveloped world. We
have analyzed an alternative approach, the Open Source model, which
is based on the idea that knowledge, unlike tangible goods, is a resource
that one can give away without being deprived of it, and therefore
should be very easy to distribute largely and equally.

By critically reviewing the essential features of Open Source software
(better termed as FLOSS) we have been able to show which ones can
be transferred to other fields of human activity, such as hardware
and intellectual work in general, giving some examples. 

Finally, we have outlined some lines of action universities can take
to spread the discussion about the Open Source model and put it in
action.

Since the Open Source model has, potentially, a revolutionary impact
on the current society, we should not expect it to spread freely and
quickly. Its many areas of conflict with the \emph{status quo} (for
example, with the publishing industry) need to be further studied
and discussed.

\subsection{A closing note}

The lesson of this study can be expressed in the simple words of a
well-known adage, that goes,
\begin{quote}
``Give a man a fish and you feed him for a day. Teach a man how to
fish and you feed him for a lifetime''
\end{quote}
We may note in closing that even the concept of ``teaching'' still
implies an asymmetry between those who hold the rights to the information
and those to whom it is administered; what is actually needed is freedom:
freedom of access and freedom of use. 

Thus we might formulate the Open Source way to sustainability as a
variant of the proverb above:
\begin{quote}
``Let \emph{every} man \emph{learn} how to fish, and you feed \emph{the
whole humanity} forever''
\end{quote}
\bibliographystyle{plainnat}
\bibliography{opensource_arx}

\end{document}